\documentclass[twocolumn,english,aps,pra,superscriptaddress]{revtex4}
\usepackage[T1]{fontenc}
\usepackage[latin9]{inputenc}
\usepackage{amsmath}
\usepackage{amssymb}
\usepackage{color}
\makeatletter

\newcommand{\ket}[1]{| #1 \rangle}
\newcommand{\bra}[1]{\langle #1 |}

\newcommand{\oper}[1]{\mathbf{\mathsf{#1}}}

\makeatother

\usepackage{babel}
\begin{document}

\title{Entropic uncertainty relations for mutually unbiased periodic coarse-grained
observables resemble their discrete counterparts}

\author{\L ukasz Rudnicki}
\email{lukasz.rudnicki@ug.edu.pl}

\selectlanguage{english}%

\affiliation{International Centre for Theory of Quantum Technologies (ICTQT),
University of Gda\'{n}sk, 80-308 Gda\'{n}sk, Poland}

\affiliation{Center for Theoretical Physics, Polish Academy of Sciences, Aleja
Lotnik{\'o}w 32/46, PL-02-668 Warsaw, Poland}

\author{Stephen P. Walborn}

\affiliation{Departamento de F\'{\i}sica, Universidad
de Concepci\'on, 160-C Concepci\'on,
Chile}

\affiliation{ANID - Millennium Science Initiative Program - Millenium Institute for Research in Optics (MIRO), Universidad de Concepci\'on,
160-C Concepci\'on, Chile}

\begin{abstract}
One of the most important and useful entropic uncertainty relations concerns a $d$ dimensional system and two mutually unbiased measurements. In such a setting, the sum of two information entropies is lower bounded by $\ln d$. It has recently been shown that projective measurements subject to operational mutual unbiasedness can also be constructed in a continuous domain, with the help of periodic coarse graining. Here we consider the whole family of R{\'e}nyi entropies applied to these discretized observables and  prove that such a scheme does also admit the entropic UR mentioned above. 
\end{abstract}
\maketitle
\section{Introduction}
Uncertainty relations are often cited as a key deviation between classical and quantum physics, describing the simultaneous unpredictability of two or more properties of a quantum system. Since the development of the concept of entropy to characterize information or the lack thereof, entropic uncertainty relations (EURs) have taken on a fundamental and useful role in quantum physics and quantum information \cite{bialynicki11,Wehner_2010,coles17,toscano18}. They can be associated to secret quantum key rates \cite{berta10,furrer11,branciard12}, and used as identifiers of quantum correlations \cite{giovannetti04,guhne04,walborn09,huang10,leach10,walborn11a,gneiting11,carvalho12,schneeloch13}, for example. Additional applications can be found in a recent review \cite{coles17}.
\par
EURs exist for systems described by either discrete variables \cite{deutsch83,maassen88,coles14,rudnicki14}, continuous variables \cite{bialynicki75,bialynicki06}, or some combination of the two \cite{bialynicki84,bialynicki85,rojas95}.  A key feature of discrete systems is that EURs for two \textit{mutually unbiased} observables give lower bounds that are a function of the dimension $d$ alone. To be more precise, two $d$-dimensional operators $\oper{X}$ and $\oper{Y}$ with all eigenstates ($i,j=0,\ldots,d-1$) satisfying $|\langle X_i | Y_j\rangle|=1/\sqrt{d}$ render two mutually unbiased measurements. For this case, an EUR involving R{\'e}nyi entropies (with natural logarithm) of orders $\alpha$ and $\beta$, such that $1/\alpha+1/\beta=2$, is given by \cite{maassen88} 
\begin{equation}
H_\alpha[X] + H_\beta[Y] \geq \ln d,
\label{eq:dsur}
\end{equation}
where 
\begin{equation} \label{Renyi0}
    H_\alpha[X]=\frac{1}{1-\alpha}\ln \sum_{i=0}^{d-1} p_i^\alpha[X],
\end{equation} and $p_i[X]=\langle X_i |\rho |X_i \rangle$. As usual, $\rho$ represents the density matrix describing the system.

\par
In the continuous-variable scenario, similar types of EURs have been developed.  However, the crucial difference between the discrete and the continuous case is that finite dimension $d$ is `lost' within a standard treatment, being replaced by a scaling parameter, related to the observables in question. To understand that effect we shall first observe that a system of mutually unbiased measurements can be characterized by two \textit{a priori} independent parameters: the number of possible measurement outcomes and the uniform 'overlap' between different measurements. The first parameter is formally the same as the number of projectors forming the resolution of the identity, and is assumed here to be the same for both measurements. The latter one has a clear operational meaning \cite{tasca18a,Tavakolieabc3847} for all projective measurements (for general POVMs it is more complicated \cite{Kalev_2014}), being equal to the true overlap between the eigenstates, in a special case of rank $1$ projectors. 

Clearly, if the first parameter is finite (therefore discrete), conservation of probability fixes the value of the latter one, as explained above Eq. \eqref{eq:dsur}. However, for continuous variables the number of outcomes is usually considered to be infinite, both countably and uncountably. As a consequence, the overlap becomes a free, setup-dependent, scaling parameter.

To see this, let us consider phase-space quadrature variables, given by $\oper{q}_\theta=\cos \theta \oper{x}+\sin \theta \oper{p}$, where    
$\oper{x}$ and $\oper{p}$, recovered for $\theta=0$ and $\theta=\pi/2$ respectively, are the usual position and momentum operators obeying $[\oper{x},\oper{p}]=i \hbar$. The commutator $[\oper{q}_\theta,\oper{q}_{\theta^\prime} ]=i\hbar \sin \Delta \theta$ clearly depends upon the relative angle between the operators  $\Delta \theta = \theta- \theta^\prime$. Therefore, uncountably many eigenstates of these operators are mutually unbiased, with overlaps given by \footnote{Proper consideration of the proper limit in the case of $\Delta \theta=0$ gives a RHS of $\delta(q_\theta-q^\prime_\theta)$.}: 
\begin{equation}\label{ovl}
|\langle q_\theta | q_{\theta^\prime}\rangle | = \left (2 \pi\hbar |\sin \Delta \theta |\right)^{-1/2}.
\end{equation}
Mutual unbiasedness of both measurements is encoded in the fact that the above overlap neither depends on $q_\theta$ nor on $q^\prime_\theta$.

\begin{table*}[t!]
\begin{centering}
\begin{tabular}{|c|c|c|}
\hline 
Number of measurements' outcomes & Overlap between the measurements & Entropic URs \tabularnewline 
\hline 
\hline 
Uncountably infinite &  $\left(2 \pi\hbar |\sin \Delta \theta |\right)^{-1/2}$ & \cite{bialynicki75,bialynicki06,huang11,guanlei09}\tabularnewline
\hline 
Countably infinite  & Additionally depends on coarse graining widths & \cite{bialynicki84,bialynicki06,rudnicki12b,rudnicki15} \tabularnewline
\hline 
Discrete, equal to $d$ & Always $1/\sqrt{d}$ & \textit{Present paper} \tabularnewline
\hline 
\end{tabular}\caption{Different types of settings relevant for continuous-variable systems and associated, known EURs. Here we fill the gap of a discrete setting.\label{Tableb1}}
\par\end{centering}
\end{table*}

In other words, the indicator of systems' dimension $d$ is replaced by $2 \pi\hbar |\sin \Delta \theta |$ --- the continuous parameter which depends on both the underlying structure of the phase space (presence of $\hbar$), and the interrelation between the involved operators, quantified by $\sin \Delta \theta$. As a natural consequence, the EURs expressed in terms of continuous R\'enyi  \cite{bialynicki06,guanlei09} and Shannon \cite{bialynicki75,huang11} entropies do depend on both parameters. We go back to these types of EURs in Sec. \ref{Sec4}. From now on we also set $\hbar\equiv 1$.

\par
An additional scaling factor arises when one takes into account that the above eigenstates describe a non-physical scenario of infinite energy, and consequently, physical scenarios involve some sort of coarse graining.  That is, the eigenstates $\ket{q_\theta}$ are approximated by ``smeared" quantum states $\int dq^\prime_\theta Q(q_\theta-q^\prime_\theta) \ket{q^\prime_\theta}$, where $Q(q_\theta-q^\prime_\theta)$ is a square integrable function that is localized around $q_\theta$ with some finite width parameter $\delta_\theta$.   Likewise, though this is just an analogy rather than a formal continuation of the previous argument, physical measurement devices (detectors) cannot be described by uncountably many rank-one projectors $\ket{q_\theta}\bra{q_\theta}$, but rather by countably many (though, still infinite number of) integrated projective measurements of the form
\begin{equation}\label{CGop}
\int_{q_\theta-\delta_\theta/2}^{q_\theta+\delta_\theta/2} dq^\prime_\theta \ket{q^\prime_\theta}\bra{q^\prime_\theta}.
\end{equation}

Adequate consideration of coarse graining in this context leads to URs with lower bounds that in addition depend explicitly on the width parameters $\delta_\theta$ \cite{bialynicki84,partovi83,bialynicki06,rudnicki12a,rudnicki12b}.  Improper attention to this inherent coarse graining can have detrimental consequences \cite{ray13a,ray13b,tasca13}. An overview of URs for coarse-grained CVs can be found in Ref. \cite{toscano18}.   

Table \ref{Tableb1} 
 summarizes the above cases. As can be seen, only settings with infinite number of outcomes have so far been successfully considered in the continuous scenario, even though, only a discrete one can lead to a counterpart of the EUR in Eq. \eqref{eq:dsur}. Therefore, as emphasized in the bottom right cell of Table \ref{Tableb1},  the aim of this paper is to provide a setting which obeys (\ref{eq:dsur}) for continuous variables. 
\par
To this end we need an alternative approach to the standard coarse graining described by (\ref{CGop}), i.e. other methods of binning together the rank-one projectors.  A number of strategies have been adopted in this direction \cite{gilchrist98,banaszek99,wenger03,vernazgris14,ketterer16,finot17}.  With the goal of defining truly mutual unbiased measurements in CV systems, periodic coarse graining (PCG) has been a successful approach.  That is, two sets of CV phase-space projectors $\Pi_k[\theta]$ and $\Pi_l[\theta']$ (like before $k,l=0,\ldots,d-1$)  can be defined such that their eigenstates give equal probability outcomes when the other measurement operator is applied \cite{tasca18a,paul18}.  This may seem to suggest that one can define a discrete variable system within a CV one, which may be loosely true, but not in any rigorous sense.  For example, it was shown that these PCG observables, though mutually unbiased, do not follow the known conditions concerning the number of allowed mutually unbiased bases for discrete systems.  Rather, depending on the number of outcomes $d$, they can mimic either the discrete or continuous cases, or neither \cite{silva21}.    
\par
Here we explore another way to benchmark PCG observables, that is, through the corresponding EURs.  We show that they indeed mimic the discrete case in that they obey the entropic URs from Eq. \eqref{eq:dsur}.  This applies to PCG of usual position and momentum operators, as well as arbitrary phase space operators. In this way, we realize our main goal, solving the problem posed in the previous paragraphs. In addition, we use our results to study the continuous limit of the entropic uncertainty relations.
    
The paper is organized as follows. In Sec. \ref{Sec2} we briefly introduce the PCG observables, while in Sec. \ref{Sec3}, we prove our major EUR for the special case of the position and momentum pair. We also state the same result for arbitrary phase-space variables. In Sec. \ref{Sec4} we study the continuous limit of the EURs considered.

\section{periodic coarse-grained observables} \label{Sec2}
In order to construct coarse-grained mutually unbiased projective measurements, we group rank one projectors according to periodic bin functions ($k=0,\dots,d-1$) \cite{tasca18b} 
\begin{equation} \label{Eq:MaskFuncDef} M_k(z;T)=\left\{ \begin{array}{ccc}   1, &\; k \,s \leq z  {\rm \,(mod \, T)}  < (k+1) s \\   0, &  {\rm otherwise}  \end{array} \right. .
\end{equation} 
The bin functions can be thought of as continuous square waves with spatial period $T$ and bin width $s=T/d$.

While for simplicity, in Sec. \ref{Sec3}, we first consider the special case of position and momentum, we now introduce notation which covers a general pair of phase-space directions. Let
\begin{equation}
\Pi_{k}[\theta]=\int_{\mathbb{R}}d q_\theta\,M_{k}\left(q_\theta;T_{\theta}\right)\left|q_\theta\right\rangle \left\langle q_\theta\right|,
\end{equation}
for $k=0,\ldots,d-1$ be a set of $d$ projectors rendering PCG in the $\theta$ direction of the phase space. In \cite{tasca18a, paul18} additional displacement parameters setting the origin of the phase space have been introduced. However, as these degrees of freedom do not at all influence the present discussion, they are omitted here. One just needs to remember that all arguments remain valid independent of the choice of the origin of the phase space. 

Given a mixed state $\rho$, we further define the probabilities
\begin{equation}\label{probs}
p_{k}[\theta]=\mathrm{Tr}\left(\rho\Pi_{k}[\theta]\right).
\end{equation}
Operational mutual unbiasedness of two measurements has been defined for pure states in \cite{tasca18a} (see also \cite{Tavakolieabc3847}), however, one can easily realize that this definition extends to the case of mixed states by convexity. To be more precise, we call both $\theta$ and $\theta'$ measurements as mutually unbiased if for all states $\rho$ such that $p_k[\theta]$ is a permutation of $(1,0,\ldots,0)$ with $d-1$ zeros, we find that   $p_l[\theta']=1/d$ for all $l$, and \textit{vice versa}. 

It is quite straightforward to realize that for $\rho=\sum_n \lambda_n \ket{\Psi_n}\bra{\Psi_n}$, with all $\lambda_n\geq 0$ and $\sum_n \lambda_n=1$, the requirement $p_k[\theta]=1$ for some $k$ enforces $\bra{\Psi_n}\Pi_k[\theta]\ket{\Psi_n}=1$ for all $n$. Consequently $\bra{\Psi_n}\Pi_l[\theta']\ket{\Psi_n}=1/d$.

Note that the above operational definition of mutual unbiasedness, as well as its natural extension to the case of mixed states, applies to any pair of projective measurements, not necessarily being the PCG, which we use here for the sake of illustration and further discussion.

In \cite{paul18} it has been proven that if
\begin{equation}
\frac{T_{\theta}T_{\theta'}}{2\pi}=\frac{d |\sin \Delta \theta |}{M},\quad M\in\mathbb{N},\quad\forall_{n=1,\ldots,d-1}\;\frac{M\,n}{d}\notin\mathbb{N},\label{condition}
\end{equation}
with $M$ being a natural number ($M\neq0$) such that $M\,n/d\notin\mathbb{N}$
for all $n=1,\ldots,d-1$ (i.e. $M$ is not co-prime with $d$), then both sets of the PCG projectors are mutually unbiased. 

\section{Entropic URs for PCG} \label{Sec3}

We are interested in an entropic UR of the general form
\begin{equation}
H_{\alpha}\left[\theta\right]+H_{\beta}\left[\theta'\right]\geq-2\ln\mathcal{C},
\end{equation}
where as usual $1/\alpha+1/\beta=2$ and the R{\'e}nyi entropy is defined in (\ref{Renyi0}).
Our aim is to show that $\mathcal{C}\leq 1/\sqrt{d}$. To this end we partially follow \cite{rudnicki10} and \cite{rudnicki12b}.

We first introduce a few pieces of notation. Let $O_{k}[\theta]$ be sets defined as
\begin{equation}
O_{k}[\theta]=\left\{ z\in\mathbb{R}:\quad M_k\left(z;T_\theta\right)=1\right\},
\end{equation}
and note that
\begin{equation}
\Pi_{k}[\theta]=\int_{O_{k}[\theta]}dq_\theta\left|q_\theta\right\rangle \left\langle q_\theta\right|.
\end{equation}

From now on we focus our attention on the position/momentum couple, further denoting $O_{k}[x]\equiv O_{k}[0]$, $O_{k}[p]\equiv O_{k}[\pi/2]$, $T_x\equiv T_0$ and $T_p\equiv T_{\pi/2}$. We define $\varphi_{km}\left(x\right)$ and $\xi_{ln}\left(p\right)$
to be orthonormal and complete sets of functions on $O_{k}[x]$ and $O_{l}[p]$
respectively, i.e.
\begin{subequations}\label{ort}
\begin{equation}
\int_{O_{k}[x]}\!\!dx\,\varphi_{k_{1}m}\left(x\right)\varphi_{k_{2}m'}^{*}\left(x\right)=\delta_{k_{1}k}\delta_{k_{2}k}\delta_{m'm},
\end{equation}
\begin{equation}
\int_{O_{l}[p]}\!\!dp\,\xi_{l_{1}n}\left(p\right)\xi_{l_{2}n'}^{*}\left(p\right)=\delta_{l_{1}l}\delta_{l_{2}l}\delta_{n'n}.
\end{equation}
\end{subequations}
Such complete sets are guaranteed to exist, since functions supported on, e.g. $O_{k}[x]$ form a subspace of the Hilbert space of square integrable functions, which is separable (so is every subspace).

Moreover, without loss of generality we restrict our attention to pure states $\rho= \ket{\Psi}\bra{\Psi}$, since  they are known to cover extreme points of information entropies. We therefore define amplitudes:
\begin{subequations}
\begin{equation}
a_{km}=\int_{O_{k}[x]}dx\,\psi\left(x\right)\varphi_{km}^{*}\left(x\right),
\end{equation}
\begin{equation}
b_{ln}=\int_{O_{l}[p]}dp\,\tilde{\psi}\left(p\right)\xi_{ln}^{*}\left(p\right),
\end{equation}
\end{subequations}
where as usual $\psi\left(x\right) = \left\langle x\left|\Psi\right\rangle \right.$ and $\tilde\psi\left(p\right) = \left\langle p\left|\Psi\right\rangle \right.\!$.

Generalizing Eqs. A7-A9 from \cite{rudnicki12b}, by replacing the intervals
appearing there by the sets $O_{k}[x]$ and $O_{l}[p]$, and with a slight adjustment of the notation concerning arguments of the R{\'e}nyi entropies, we immediately
get the result
\begin{equation}\label{Ren1}
H_{\alpha}\left[\left|a\right|^{2}\right]+H_{\beta}\left[\left|b\right|^{2}\right]\geq-2\ln\mathcal{C},
\end{equation}
where
\begin{equation}\label{Cc}
\mathcal{C}=\sup_{\left(k,l,m,n\right)}\left|\int_{O_{k}[x]}dx\int_{O_{l}[p]}dp\frac{e^{ipx}}{\sqrt{2\pi}}\varphi_{km}^{*}\left(x\right)\xi_{ln}\left(p\right)\right|.
\end{equation}
Arguments of the R{\'e}nyi entropies in (\ref{Ren1}) are not denoted as the directions on the phase space, but as the probability distributions entering Eq. \eqref{Renyi0}.
These distributions are more fine-grained than (\ref{probs}), since 
\begin{equation}
    p_k[0]=\sum_m \left|a_{km}\right|^{2},\quad p_l[\pi/2]=\sum_n \left|b_{ln}\right|^{2}.
\end{equation}

If we further apply the Cauchy-Schwarz inequality to the $\int_{O_{k}[x]}dx$
integral and use normalization of $\varphi_{km}\left(x\right)$, 
we arrive at the bound $\mathcal{C}\leq \sup_{\left(k,l,n\right)} W_{kl}^n $, where
\begin{equation}
W_{kl}^n=\sqrt{\int_{O_{k}[x]}\!\!\!\!dx\int_{O_{l}[p]}\!\!\!\!dp\int_{O_{l}[p]}\!\!\!\!dp'\frac{e^{i\left(p-p'\right)x}}{2\pi}\xi_{ln}\left(p\right)\xi_{ln}^{*}\left(p'\right)}.
\label{eq:Csq}
\end{equation}
Our task is therefore to compute the kernel
\begin{equation}\label{kernel}
\int_{O_{k}[x]}dx\frac{e^{i\left(p-p'\right)x}}{2\pi}.
\end{equation}

The periodic bin function can be decomposed in the Fourier series
\begin{equation}
M_{k}\left(z;T\right)=\frac{1}{d}+\!\!\sum_{N\in\mathbb{Z}/\{0\}}\!\!f_{k,N}e^{\frac{2\pi iN}{T}z},
\label{eq:M}
\end{equation}
where 
\begin{equation}\label{fn}
f_{k,N}=\frac{1-e^{-\frac{2\pi iN}{d}}}{2\pi iN}e^{-\frac{2\pi iN}{d}k}.
\end{equation}
Using \eqref{eq:M} to calculate the kernel we find
\begin{widetext}
\begin{equation}\label{major1}
\int_{O_{k}[x]}dx\frac{e^{i\left(p-p'\right)x}}{2\pi}  = \frac{1}{d}\delta\left(p-p'\right) +  \!\!\sum_{N\in\mathbb{Z}/\{0\}}\!\!f_{k,N}\delta\left(p-p'+\frac{2\pi}{T_{x}}N\right).
\end{equation}
Consequently, we obtain the result 
\begin{equation}\label{major2}
\int_{O_{k}[x]}\!\!\!\!dx\int_{O_{l}[p]}\!\!\!\!dp\int_{O_{l}[p]}\!\!\!\!dp'\frac{e^{i\left(p-p'\right)x}}{2\pi}\xi_{ln}\left(p\right)\xi_{ln}^{*}\left(p'\right)=\frac{1}{d}+\!\!\sum_{N\in\mathbb{Z}/\{0\}}\!\!f_{k,N}\int_{O_{l}[p]}\!\!\!\!dp\int_{O_{l}[p]}\!\!\!\!dp'\delta\left(p-p'+\frac{T_{p}MN}{d}\right)\xi_{ln}\left(p\right)\xi_{ln}^{*}\left(p'\right),
\end{equation}
\end{widetext} where we have utilized normalization of $\xi_{ln}\left(p\right)$ to integrate the first Dirac delta contribution, 
and we applied the MUB condition (\ref{condition}) while changing arguments of the remaining Dirac deltas. Due to the last step, every Dirac delta in the second
expression leads to an autocorrelation term, which is non-vanishing only
when $MN/d$ is an integer. However, due to the further requirement established in (\ref{condition}), we find
$MN/d\in\mathbb{Z}$ if and only if $N/d\in\mathbb{Z}$. But in this
special case the factor $1-e^{-\frac{2\pi iN}{d}}$ present in $f_{k,N}$ becomes equal
to $0$, so that all terms in the sum over $N\in\mathbb{Z}/\{0\}$
disappear, leaving the bare contribution $1/d$. As a result, $\mathcal{C}\leq1/\sqrt{d}$, as expected. 

Finally, we observe \cite{rudnicki12b} that a particular choice 
\begin{subequations}\label{prob}
\begin{equation}
\varphi_{k0}\left(x\right)=\left\langle x\right|\Pi_{k}[0]\left|\Psi\right\rangle /\sqrt{p_{k}[0]},
\end{equation}
\begin{equation}
\xi_{l0}\left(p\right)=\left\langle p\right|\Pi_{l}[\pi/2]\left|\Psi\right\rangle /\sqrt{p_{l}[\pi/2]},
\end{equation}
\end{subequations}
with other functions in both complete sets being orthogonal to (\ref{prob}) leads to the probabilities $\left|a_{km}\right|^{2}=p_{k}[0]\delta_{m0}$
and $\left|b_{ln}\right|^{2}=p_{l}[\pi/2]\delta_{n0}$. Therefore, the bound
$\mathcal{C}\leq1/\sqrt{d}$ which consequently gives $-2\ln\mathcal{C}\geq\ln d$
is also valid for our main UR under consideration, namely, Eq. (\ref{eq:dsur}) for position and momentum pair of PCG observables, denoted by angles $\theta=0$ and $\theta=\pi/2$ respectively, is proven. It is easy to recognize that this bound, due to the property
of mutual unbiasedness, is saturated if a state is localized in either
of the sets $O_{k}[x]$ or $O_{l}[p]$, for a fixed value of the index $k$
or $l$.

\subsection{Extension to any two directions in phase space}
In order to extend the above result to two arbitrary phase-space observables $q_\theta$ and $q_{\theta^\prime}$, i.e. to show that the general EUR
\begin{equation}
H_{\alpha}\left[\theta\right]+H_{\beta}\left[\theta^\prime\right]\geq\ln d,\label{eq:main2}
\end{equation}
\medskip
holds (as always with $1/\alpha +1/\beta=2$), we first observe that several steps of the previous derivation can immediately be repeated with minor modifications. To be more precise, Eqs. (\ref{ort})-(\ref{fn}) from Sec. \ref{Sec3} just require a slight adjustment of the notation, which boils down to a replacement of labels "$0$" or "$x$" by "$\theta$" and  "$\pi/2$" or "$p$" by "$\theta^\prime$", as well as function arguments by  $q_\theta$ and $q_\theta^\prime$, respectively. Moreover, the Fourier transform in (\ref{Cc}) must be replaced by the fractional Fourier transform \cite{FFT} which gives the generalized overlap 
$\left\langle q_{\theta}\left|q_{\theta^\prime}\right\rangle \right.\!=\mathcal{F}\left(q_{\theta},q_{\theta^\prime}\right)$ and reads (as before $\Delta\theta=\theta-\theta^\prime$)
\begin{equation}
  \mathcal{F}\left(q_{\theta},q_{\theta^\prime}\right)=\sqrt{\frac{-ie^{i\Delta\theta}}{2\pi\sin\Delta\theta}}e^{i\frac{\cot\Delta\theta}{2}\left(q_{\theta}^{2}+q_{\theta'}^{2}\right)-i\frac{q_{\theta}q_{\theta'}}{\sin\Delta\theta}}.
\end{equation}
Note that $\left|  \mathcal{F}\left(q_{\theta},q_{\theta^\prime}\right)\right|$ reduces to the overlap in (\ref{ovl}).
Consequently, the Fourier kernel (\ref{kernel}) is replaced by
\begin{equation}
\int_{O_{k}[\theta]}dq_\theta\mathcal{F}\left(q_{\theta},q_{\theta^\prime}\right)\mathcal{F}\left(\tilde{q}_{\theta^\prime},q_\theta\right).
\end{equation}
Since 
\begin{equation}
    \mathcal{F}\left(q_{\theta},q_{\theta^\prime}\right)\mathcal{F}\left(\tilde{q}_{\theta^\prime},q_\theta\right)=\frac{e^{i\frac{\cot\Delta\theta}{2}\left(q_{\theta'}^{2}-\tilde{q}_{\theta'}^{2}\right)}}{2\pi\left|\sin\Delta\theta\right|}e^{i\frac{q_{\theta}}{\sin\Delta\theta}\left(\tilde{q}_{\theta'}-q_{\theta'}\right)},
\end{equation}
we easily generalize (\ref{major1}) as
\begin{widetext}
\begin{equation}
\int_{O_{k}[\theta]}dq_\theta\mathcal{F}\left(q_{\theta},q_{\theta^\prime}\right)\mathcal{F}\left(\tilde{q}_{\theta^\prime},q_\theta\right) = \frac{1}{d}\delta\left(\tilde{q}_{\theta'}-q_{\theta'}\right) +  \!\!\sum_{N\in\mathbb{Z}/\{0\}}\!\!f_{k,N} e^{i\frac{\cot\Delta\theta}{2}\left(q_{\theta'}^{2}-\tilde{q}_{\theta'}^{2}\right)}\delta\left(\tilde{q}_{\theta'}-q_{\theta'}+\frac{2\pi\sin\Delta\theta}{T_{\theta}}N\right).
\end{equation}
\end{widetext}
The remaining part of the derivation follows exactly the same way as for the particular case of position and momentum. The only difference is that due to the MUB condition Eq. \eqref{condition}, the term $2\pi\sin\Delta\theta/T_{\theta}$ inside the Dirac delta is replaced by $\pm\, T_{\theta^\prime} M/d$ where the sign $\pm$ depends on the order of $\theta$ and $\theta^\prime$ on the phase space. In (\ref{major2}) we find the plus sign as the angle difference for position and momentum is in $[0,\pi]$. Thus, we have an uncertainty relation of the form (\ref{eq:dsur}) for PCG observables corresponding to any two non-parallel phase space quadratures.

\section{Continuous limit} \label{Sec4}
At the end we would briefly like to elaborate on the continuous limit for PCG observables.  To this end we recall that $d=T_\theta/s_\theta$, for all variables $q_\theta$, where $s_\theta$ is the bin width.   Then, using \eqref{condition}, we can write $d = 2 \pi |\sin \Delta \theta|/s_\theta s_{\theta^\prime}$.  Plugging this into \eqref{eq:main2}, we have 
\begin{equation}
H_{\alpha}\left[\theta\right] +H_{\beta}\left[\theta^\prime\right] + \ln(s_\theta s_{\theta^\prime}) \geq\ln 2 \pi |\sin \Delta \theta| .
\label{eq:dur}
\end{equation}
Each R{\'e}nyi entropy can be rewritten as follows
\begin{equation}\label{RenCont}
    H_{\alpha}\left[\theta\right]=-\ln s_\theta+\frac{1}{1-\alpha}\ln\left[ \sum_{i=0}^{d-1}s_\theta \left(\frac{p_i[\theta]}{s_\theta}\right)^\alpha\right].
\end{equation}

 In the continuous limit $d\rightarrow \infty$, we set $T_{\theta}\sim\sqrt{d}$, so that $T_{\theta}\rightarrow \infty$ while at the same time $s_\theta\rightarrow 0$. In this limit, the sum multiplied by $s_\theta$ tends to the integral $\int_0^\infty d q_\theta$, while the term in parenthesis in (\ref{RenCont}) becomes a continuous probability distribution supported on $[0,\infty)$. This specific probability distribution takes into account two points on the real line, one on the positive side and one the negative side (though not symmetrically). To explain it a bit better we can for the moment restrict ourselves to a box $[-L,L]$ and, given a function $f(x)$ supported on that box, consider the function $g(x)=f(x)+f(x-L)$, which is supported on $[0,L]$. In our limiting procedure, the continuous probability distributions on the real line, which normally are the arguments of the R{\'e}nyi entropies, will be of the $f$-type, while $p_i[\theta]/s_\theta$ tends to the distribution of the $g$-type. Obviously, the $g$-type probability distributions will always have smaller entropy than $f$-type distributions. Therefore, the continuous R{\'e}nyi entropy $h_{\alpha}\left[\theta\right]$ will also be bigger than the continuous limit of the entropy in (\ref{RenCont})
 \begin{equation}
     h_{\alpha}\left[\theta\right]\geq \lim_{d\rightarrow \infty} \left(H_{\alpha}\left[\theta\right]+\ln s_\theta\right).
 \end{equation}
 As a consequence, using \eqref{eq:dur}, we obtain the continuous UR
 \begin{equation}
h_{\alpha}\left[\theta\right] +h_{\beta}\left[\theta^\prime\right]  \geq\ln 2 \pi |\sin \Delta \theta|.
\end{equation}
The uncertainty relation obtained is clearly weaker than the best known URs for continuous variables \cite{bialynicki06,guanlei09}. This is because the latter follow from a completely different mathematical machinery, namely "$p$-$q$ norm" inequalities for the Fourier transform. Our result is on the contrary closest in spirit with standard finite-dimensional treatment of mutually unbiased bases which, while powerful, does not know much about sophisticated properties of the Fourier transform.

\section{Discussion}
We have provided an entropic uncertainty relation for a discrete set of mutually unbiased, periodic coarse-grained observables.  Different from the underlying continuous observables, or other discretization schemes, here the uncertainty limit is bounded only by the number of measurement outcomes, which plays the role of dimension.    
We extend our results to apply to observables constructed from eigenstates of any two non-parallel quadrature operators, and show that a meaningful (though not optimal) continuous limit can be obtained.  
\par
A number of possible applications and open questions exist.  First, it is tempting to ask whether these results can be extended to include more than two observables, as in  \cite{paul18,silva21}. This remains an open question, since to date an entropic uncertainty relation for more than two continuous operators has not been proven \cite{weigert08}, and the established results for discrete systems \cite{SANCHEZ1993233,PhysRevA.79.022104,PhysRevA.92.032109} do not seem to be directly applicable. As an application of our results, the EUR derived here for PCG observables can be  adapted to identify entanglement, or more specifically, as a criteria for EPR-steering correlations \cite{wiseman07} between two parties.  For example, it is straightforward to follow the recipe in Refs. \cite{walborn11a,schneeloch13}, which, for $\alpha=\beta=1$ (Shannon entropies), leads to 
\begin{equation}
H_{1}\left[q_\theta|q_\phi\right] +H_{1}\left[q_{\theta^\prime}|q_{\phi^\prime}\right] \geq\ln d,
\end{equation}
where $H_{1}\left[r|s\right]$ is the conditional Shannon entropy and $r$ and $s$ refer to measurement directions of Alice and Bob (the two parties). Violation of the above inequality indicates EPR steering correlations in Alice and Bob's bipartite system.
\par
The main motivation for our work is the overall question concerning the behaviour of discretized observables constructed within a continuous Hilbert space, and whether these observables are ``more continuous" or ``more discrete" in their characteristics.  Our results show in the case of periodic coarse graining, the discretization is indeed manifest in the desired way, whereas entropic uncertainty relations are applied.

\acknowledgments 
\L .R. acknowledges support by the Foundation for Polish Science (IRAP
project, ICTQT, Contract No. 2018/MAB/5, cofinanced by the EU within
the Smart Growth Operational Programme).  SPW was supported by Fondo Nacional de Desarrollo Cient\'{i}fico y Tecnol\'{o}gico (ANID) (1200266) and  ANID - Millennium Science Initiative Program  -  ICN17\_012.

\bibliographystyle{apsrev}

\begin{thebibliography}{53}
\expandafter\ifx\csname natexlab\endcsname\relax\def\natexlab#1{#1}\fi
\expandafter\ifx\csname bibnamefont\endcsname\relax
  \def\bibnamefont#1{#1}\fi
\expandafter\ifx\csname bibfnamefont\endcsname\relax
  \def\bibfnamefont#1{#1}\fi
\expandafter\ifx\csname citenamefont\endcsname\relax
  \def\citenamefont#1{#1}\fi
\expandafter\ifx\csname url\endcsname\relax
  \def\url#1{\texttt{#1}}\fi
\expandafter\ifx\csname urlprefix\endcsname\relax\def\urlprefix{URL }\fi
\providecommand{\bibinfo}[2]{#2}
\providecommand{\eprint}[2][]{\url{#2}}

\bibitem[{\citenamefont{Bialynicki-Birula and Rudnicki}(2011)}]{bialynicki11}
\bibinfo{author}{\bibfnamefont{I.}~\bibnamefont{Bialynicki-Birula}}
  \bibnamefont{and}
  \bibinfo{author}{\bibfnamefont{{\L}.}~\bibnamefont{Rudnicki}},
  \emph{\bibinfo{title}{Entropic Uncertainty Relations in Quantum Physics}}
  (\bibinfo{publisher}{Springer}, \bibinfo{address}{Dordrecht},
  \bibinfo{year}{2011}), p.~\bibinfo{pages}{1}.

\bibitem[{\citenamefont{Wehner and Winter}(2010)}]{Wehner_2010}
\bibinfo{author}{\bibfnamefont{S.}~\bibnamefont{Wehner}} \bibnamefont{and}
  \bibinfo{author}{\bibfnamefont{A.}~\bibnamefont{Winter}},
  \bibinfo{journal}{New Journal of Physics} \textbf{\bibinfo{volume}{12}},
  \bibinfo{pages}{025009} (\bibinfo{year}{2010}),
  \urlprefix\url{https://doi.org/10.1088/1367-2630/12/2/025009}.

\bibitem[{\citenamefont{Coles et~al.}(2017)\citenamefont{Coles, Berta,
  Tomamichel, and Wehner}}]{coles17}
\bibinfo{author}{\bibfnamefont{P.~J.} \bibnamefont{Coles}},
  \bibinfo{author}{\bibfnamefont{M.}~\bibnamefont{Berta}},
  \bibinfo{author}{\bibfnamefont{M.}~\bibnamefont{Tomamichel}},
  \bibnamefont{and} \bibinfo{author}{\bibfnamefont{S.}~\bibnamefont{Wehner}},
  \bibinfo{journal}{Rev. Mod. Phys.} \textbf{\bibinfo{volume}{89}},
  \bibinfo{pages}{015002} (\bibinfo{year}{2017}),
  \urlprefix\url{https://link.aps.org/doi/10.1103/RevModPhys.89.015002}.

\bibitem[{\citenamefont{Toscano et~al.}(2018)\citenamefont{Toscano, Tasca,
  Rudnicki, and Walborn}}]{toscano18}
\bibinfo{author}{\bibfnamefont{F.}~\bibnamefont{Toscano}},
  \bibinfo{author}{\bibfnamefont{D.~S.} \bibnamefont{Tasca}},
  \bibinfo{author}{\bibfnamefont{{\L}.}~\bibnamefont{Rudnicki}},
  \bibnamefont{and} \bibinfo{author}{\bibfnamefont{S.~P.}
  \bibnamefont{Walborn}}, \bibinfo{journal}{Entropy}
  \textbf{\bibinfo{volume}{20}}, \bibinfo{pages}{454} (\bibinfo{year}{2018}).

\bibitem[{\citenamefont{Berta et~al.}(2010)\citenamefont{Berta, Christandl,
  Colbeck, Renes, and Renner}}]{berta10}
\bibinfo{author}{\bibfnamefont{M.}~\bibnamefont{Berta}},
  \bibinfo{author}{\bibfnamefont{M.}~\bibnamefont{Christandl}},
  \bibinfo{author}{\bibfnamefont{R.}~\bibnamefont{Colbeck}},
  \bibinfo{author}{\bibfnamefont{J.~M.} \bibnamefont{Renes}}, \bibnamefont{and}
  \bibinfo{author}{\bibfnamefont{R.}~\bibnamefont{Renner}},
  \bibinfo{journal}{Nat. Phys.} \textbf{\bibinfo{volume}{6}},
  \bibinfo{pages}{659} (\bibinfo{year}{2010}).

\bibitem[{\citenamefont{Furrer et~al.}(2011)\citenamefont{Furrer, {\AA}berg,
  and Renner}}]{furrer11}
\bibinfo{author}{\bibfnamefont{F.}~\bibnamefont{Furrer}},
  \bibinfo{author}{\bibfnamefont{J.}~\bibnamefont{{\AA}berg}},
  \bibnamefont{and} \bibinfo{author}{\bibfnamefont{R.}~\bibnamefont{Renner}},
  \bibinfo{journal}{Communications in Mathematical Physics}
  \textbf{\bibinfo{volume}{306}}, \bibinfo{pages}{165} (\bibinfo{year}{2011}),
  \urlprefix\url{https://doi.org/10.1007/s00220-011-1282-1}.

\bibitem[{\citenamefont{Branciard et~al.}(2012)\citenamefont{Branciard,
  Cavalcanti, Walborn, Scarani, and Wiseman}}]{branciard12}
\bibinfo{author}{\bibfnamefont{C.}~\bibnamefont{Branciard}},
  \bibinfo{author}{\bibfnamefont{E.~G.} \bibnamefont{Cavalcanti}},
  \bibinfo{author}{\bibfnamefont{S.~P.} \bibnamefont{Walborn}},
  \bibinfo{author}{\bibfnamefont{V.}~\bibnamefont{Scarani}}, \bibnamefont{and}
  \bibinfo{author}{\bibfnamefont{H.~M.} \bibnamefont{Wiseman}},
  \bibinfo{journal}{Phys. Rev. A} \textbf{\bibinfo{volume}{85}},
  \bibinfo{pages}{010301} (\bibinfo{year}{2012}).

\bibitem[{\citenamefont{Giovannetti et~al.}(2004)\citenamefont{Giovannetti,
  Lloyd, and Maccone}}]{giovannetti04}
\bibinfo{author}{\bibfnamefont{V.}~\bibnamefont{Giovannetti}},
  \bibinfo{author}{\bibfnamefont{S.}~\bibnamefont{Lloyd}}, \bibnamefont{and}
  \bibinfo{author}{\bibfnamefont{L.}~\bibnamefont{Maccone}},
  \bibinfo{journal}{Science} \textbf{\bibinfo{volume}{306}},
  \bibinfo{pages}{1330} (\bibinfo{year}{2004}).

\bibitem[{\citenamefont{G\"uhne and Lewenstein}(2004)}]{guhne04}
\bibinfo{author}{\bibfnamefont{O.}~\bibnamefont{G\"uhne}} \bibnamefont{and}
  \bibinfo{author}{\bibfnamefont{M.}~\bibnamefont{Lewenstein}},
  \bibinfo{journal}{Phys. Rev. A} \textbf{\bibinfo{volume}{70}},
  \bibinfo{pages}{022316} (\bibinfo{year}{2004}),
  \urlprefix\url{https://link.aps.org/doi/10.1103/PhysRevA.70.022316}.

\bibitem[{\citenamefont{Walborn et~al.}(2009)\citenamefont{Walborn, Taketani,
  Salles, Toscano, and de~Matos~Filho}}]{walborn09}
\bibinfo{author}{\bibfnamefont{S.~P.} \bibnamefont{Walborn}},
  \bibinfo{author}{\bibfnamefont{B.~G.} \bibnamefont{Taketani}},
  \bibinfo{author}{\bibfnamefont{A.}~\bibnamefont{Salles}},
  \bibinfo{author}{\bibfnamefont{F.}~\bibnamefont{Toscano}}, \bibnamefont{and}
  \bibinfo{author}{\bibfnamefont{R.~L.} \bibnamefont{de~Matos~Filho}},
  \bibinfo{journal}{Phys. Rev. Lett.} \textbf{\bibinfo{volume}{103}},
  \bibinfo{pages}{160505} (\bibinfo{year}{2009}).

\bibitem[{\citenamefont{Huang}(2010)}]{huang10}
\bibinfo{author}{\bibfnamefont{Y.}~\bibnamefont{Huang}},
  \bibinfo{journal}{Phys. Rev. A} \textbf{\bibinfo{volume}{82}},
  \bibinfo{pages}{012335} (\bibinfo{year}{2010}),
  \urlprefix\url{https://link.aps.org/doi/10.1103/PhysRevA.82.012335}.

\bibitem[{\citenamefont{Leach et~al.}(2010)\citenamefont{Leach, Jack, Romero,
  Jha, Yao, Franke-Arnold, Ireland, Boyd, Barnett, and Padgett}}]{leach10}
\bibinfo{author}{\bibfnamefont{J.}~\bibnamefont{Leach}},
  \bibinfo{author}{\bibfnamefont{B.}~\bibnamefont{Jack}},
  \bibinfo{author}{\bibfnamefont{J.}~\bibnamefont{Romero}},
  \bibinfo{author}{\bibfnamefont{A.~K.} \bibnamefont{Jha}},
  \bibinfo{author}{\bibfnamefont{A.~M.} \bibnamefont{Yao}},
  \bibinfo{author}{\bibfnamefont{S.}~\bibnamefont{Franke-Arnold}},
  \bibinfo{author}{\bibfnamefont{D.~G.} \bibnamefont{Ireland}},
  \bibinfo{author}{\bibfnamefont{R.~W.} \bibnamefont{Boyd}},
  \bibinfo{author}{\bibfnamefont{S.~M.} \bibnamefont{Barnett}},
  \bibnamefont{and} \bibinfo{author}{\bibfnamefont{M.~J.}
  \bibnamefont{Padgett}}, \bibinfo{journal}{Science}
  \textbf{\bibinfo{volume}{329}}, \bibinfo{pages}{662} (\bibinfo{year}{2010}).

\bibitem[{\citenamefont{Walborn et~al.}(2011)\citenamefont{Walborn, Salles,
  Gomes, Toscano, and Souto~Ribeiro}}]{walborn11a}
\bibinfo{author}{\bibfnamefont{S.~P.} \bibnamefont{Walborn}},
  \bibinfo{author}{\bibfnamefont{A.}~\bibnamefont{Salles}},
  \bibinfo{author}{\bibfnamefont{R.~M.} \bibnamefont{Gomes}},
  \bibinfo{author}{\bibfnamefont{F.}~\bibnamefont{Toscano}}, \bibnamefont{and}
  \bibinfo{author}{\bibfnamefont{P.~H.} \bibnamefont{Souto~Ribeiro}},
  \bibinfo{journal}{Phys. Rev. Lett.} \textbf{\bibinfo{volume}{106}},
  \bibinfo{pages}{130402} (\bibinfo{year}{2011}).

\bibitem[{\citenamefont{Gneiting and Hornberger}(2011)}]{gneiting11}
\bibinfo{author}{\bibfnamefont{C.}~\bibnamefont{Gneiting}} \bibnamefont{and}
  \bibinfo{author}{\bibfnamefont{K.}~\bibnamefont{Hornberger}},
  \bibinfo{journal}{Phys. Rev. Lett.} \textbf{\bibinfo{volume}{106}},
  \bibinfo{pages}{210501} (\bibinfo{year}{2011}).

\bibitem[{\citenamefont{Carvalho et~al.}(2012)\citenamefont{Carvalho, Ferraz,
  Borges, de~Assis, P\'adua, and Walborn}}]{carvalho12}
\bibinfo{author}{\bibfnamefont{M.~A.~D.} \bibnamefont{Carvalho}},
  \bibinfo{author}{\bibfnamefont{J.}~\bibnamefont{Ferraz}},
  \bibinfo{author}{\bibfnamefont{G.~F.} \bibnamefont{Borges}},
  \bibinfo{author}{\bibfnamefont{P.-L.} \bibnamefont{de~Assis}},
  \bibinfo{author}{\bibfnamefont{S.}~\bibnamefont{P\'adua}}, \bibnamefont{and}
  \bibinfo{author}{\bibfnamefont{S.~P.} \bibnamefont{Walborn}},
  \bibinfo{journal}{Phys. Rev. A} \textbf{\bibinfo{volume}{86}},
  \bibinfo{pages}{032332} (\bibinfo{year}{2012}).

\bibitem[{\citenamefont{Schneeloch et~al.}(2013)\citenamefont{Schneeloch,
  Dixon, Howland, Broadbent, and Howell}}]{schneeloch13}
\bibinfo{author}{\bibfnamefont{J.}~\bibnamefont{Schneeloch}},
  \bibinfo{author}{\bibfnamefont{P.~B.} \bibnamefont{Dixon}},
  \bibinfo{author}{\bibfnamefont{G.~A.} \bibnamefont{Howland}},
  \bibinfo{author}{\bibfnamefont{C.~J.} \bibnamefont{Broadbent}},
  \bibnamefont{and} \bibinfo{author}{\bibfnamefont{J.~C.}
  \bibnamefont{Howell}}, \bibinfo{journal}{Phys. Rev. Lett.}
  \textbf{\bibinfo{volume}{110}}, \bibinfo{pages}{130407}
  (\bibinfo{year}{2013}).

\bibitem[{\citenamefont{Deutsch}(1983)}]{deutsch83}
\bibinfo{author}{\bibfnamefont{D.}~\bibnamefont{Deutsch}},
  \bibinfo{journal}{Phys. Rev. Lett.} \textbf{\bibinfo{volume}{50}},
  \bibinfo{pages}{631} (\bibinfo{year}{1983}).

\bibitem[{\citenamefont{Maassen and Uffink}(1988)}]{maassen88}
\bibinfo{author}{\bibfnamefont{H.}~\bibnamefont{Maassen}} \bibnamefont{and}
  \bibinfo{author}{\bibfnamefont{J.~B.~M.} \bibnamefont{Uffink}},
  \bibinfo{journal}{Phys. Rev. Lett.} \textbf{\bibinfo{volume}{60}},
  \bibinfo{pages}{1103} (\bibinfo{year}{1988}).

\bibitem[{\citenamefont{Coles and Piani}(2014)}]{coles14}
\bibinfo{author}{\bibfnamefont{P.~J.} \bibnamefont{Coles}} \bibnamefont{and}
  \bibinfo{author}{\bibfnamefont{M.}~\bibnamefont{Piani}},
  \bibinfo{journal}{Phys. Rev. A} \textbf{\bibinfo{volume}{89}},
  \bibinfo{pages}{022112} (\bibinfo{year}{2014}),
  \urlprefix\url{https://link.aps.org/doi/10.1103/PhysRevA.89.022112}.

\bibitem[{\citenamefont{Rudnicki et~al.}(2014)\citenamefont{Rudnicki,
  Pucha\l{}a, and \ifmmode~\dot{Z}\else \.{Z}\fi{}yczkowski}}]{rudnicki14}
\bibinfo{author}{\bibfnamefont{{\L}.}~\bibnamefont{Rudnicki}},
  \bibinfo{author}{\bibfnamefont{Z.}~\bibnamefont{Pucha\l{}a}},
  \bibnamefont{and}
  \bibinfo{author}{\bibfnamefont{K.}~\bibnamefont{\ifmmode~\dot{Z}\else
  \.{Z}\fi{}yczkowski}}, \bibinfo{journal}{Phys. Rev. A}
  \textbf{\bibinfo{volume}{89}}, \bibinfo{pages}{052115}
  (\bibinfo{year}{2014}),
  \urlprefix\url{https://link.aps.org/doi/10.1103/PhysRevA.89.052115}.

\bibitem[{\citenamefont{Bialynicki-Birula and Mycielski}(1975)}]{bialynicki75}
\bibinfo{author}{\bibfnamefont{I.}~\bibnamefont{Bialynicki-Birula}}
  \bibnamefont{and}
  \bibinfo{author}{\bibfnamefont{J.}~\bibnamefont{Mycielski}},
  \bibinfo{journal}{Commun. Math. Phys.} \textbf{\bibinfo{volume}{44}},
  \bibinfo{pages}{129} (\bibinfo{year}{1975}).

\bibitem[{\citenamefont{Bialynicki-Birula}(2006)}]{bialynicki06}
\bibinfo{author}{\bibfnamefont{I.}~\bibnamefont{Bialynicki-Birula}},
  \bibinfo{journal}{Phys. Rev. A} \textbf{\bibinfo{volume}{74}},
  \bibinfo{eid}{052101} (\bibinfo{year}{2006}).

\bibitem[{\citenamefont{Bialynicki-Birula}(1984)}]{bialynicki84}
\bibinfo{author}{\bibfnamefont{I.}~\bibnamefont{Bialynicki-Birula}},
  \bibinfo{journal}{Phys. Lett.} \textbf{\bibinfo{volume}{103 A}},
  \bibinfo{pages}{253} (\bibinfo{year}{1984}).

\bibitem[{\citenamefont{Bialynicki-Birula and Madajczyk}(1985)}]{bialynicki85}
\bibinfo{author}{\bibfnamefont{I.}~\bibnamefont{Bialynicki-Birula}}
  \bibnamefont{and} \bibinfo{author}{\bibfnamefont{J.~L.}
  \bibnamefont{Madajczyk}}, \bibinfo{journal}{Phys. Lett.}
  \textbf{\bibinfo{volume}{108 A}}, \bibinfo{pages}{384}
  (\bibinfo{year}{1985}).

\bibitem[{\citenamefont{{Rojas Gonz{\'a}lez} et~al.}(1995)\citenamefont{{Rojas
  Gonz{\'a}lez}, Vaccaro, and Barnett}}]{rojas95}
\bibinfo{author}{\bibfnamefont{A.}~\bibnamefont{{Rojas Gonz{\'a}lez}}},
  \bibinfo{author}{\bibfnamefont{J.~A.} \bibnamefont{Vaccaro}},
  \bibnamefont{and} \bibinfo{author}{\bibfnamefont{S.~M.}
  \bibnamefont{Barnett}}, \bibinfo{journal}{Physics Letters A}
  \textbf{\bibinfo{volume}{205}}, \bibinfo{pages}{247} (\bibinfo{year}{1995}),
  ISSN \bibinfo{issn}{0375-9601},
  \urlprefix\url{https://www.sciencedirect.com/science/article/pii/037596019500582N}.

\bibitem[{\citenamefont{Tasca et~al.}(2018{\natexlab{a}})\citenamefont{Tasca,
  S\'anchez, Walborn, and Rudnicki}}]{tasca18a}
\bibinfo{author}{\bibfnamefont{D.~S.} \bibnamefont{Tasca}},
  \bibinfo{author}{\bibfnamefont{P.}~\bibnamefont{S\'anchez}},
  \bibinfo{author}{\bibfnamefont{S.~P.} \bibnamefont{Walborn}},
  \bibnamefont{and}
  \bibinfo{author}{\bibfnamefont{{\L}.}~\bibnamefont{Rudnicki}},
  \bibinfo{journal}{Phys. Rev. Lett.} \textbf{\bibinfo{volume}{120}},
  \bibinfo{pages}{040403} (\bibinfo{year}{2018}{\natexlab{a}}),
  \urlprefix\url{https://link.aps.org/doi/10.1103/PhysRevLett.120.040403}.

\bibitem[{\citenamefont{Tavakoli et~al.}(2021)\citenamefont{Tavakoli, Farkas,
  Rosset, Bancal, and Kaniewski}}]{Tavakolieabc3847}
\bibinfo{author}{\bibfnamefont{A.}~\bibnamefont{Tavakoli}},
  \bibinfo{author}{\bibfnamefont{M.}~\bibnamefont{Farkas}},
  \bibinfo{author}{\bibfnamefont{D.}~\bibnamefont{Rosset}},
  \bibinfo{author}{\bibfnamefont{J.-D.} \bibnamefont{Bancal}},
  \bibnamefont{and}
  \bibinfo{author}{\bibfnamefont{J.}~\bibnamefont{Kaniewski}},
  \bibinfo{journal}{Science Advances} \textbf{\bibinfo{volume}{7}}
  (\bibinfo{year}{2021}).

\bibitem[{\citenamefont{Kalev and Gour}(2014)}]{Kalev_2014}
\bibinfo{author}{\bibfnamefont{A.}~\bibnamefont{Kalev}} \bibnamefont{and}
  \bibinfo{author}{\bibfnamefont{G.}~\bibnamefont{Gour}}, \bibinfo{journal}{New
  Journal of Physics} \textbf{\bibinfo{volume}{16}}, \bibinfo{pages}{053038}
  (\bibinfo{year}{2014}),
  \urlprefix\url{https://doi.org/10.1088/1367-2630/16/5/053038}.

\bibitem[{\citenamefont{Huang}(2011)}]{huang11}
\bibinfo{author}{\bibfnamefont{Y.}~\bibnamefont{Huang}},
  \bibinfo{journal}{Phys. Rev. A} \textbf{\bibinfo{volume}{83}},
  \bibinfo{pages}{052124} (\bibinfo{year}{2011}).

\bibitem[{\citenamefont{Guanlei et~al.}(2009)\citenamefont{Guanlei, Xiaotong,
  and Xiaogang}}]{guanlei09}
\bibinfo{author}{\bibfnamefont{X.}~\bibnamefont{Guanlei}},
  \bibinfo{author}{\bibfnamefont{W.}~\bibnamefont{Xiaotong}}, \bibnamefont{and}
  \bibinfo{author}{\bibfnamefont{X.}~\bibnamefont{Xiaogang}},
  \bibinfo{journal}{Signal Processing} \textbf{\bibinfo{volume}{89}},
  \bibinfo{pages}{2692 } (\bibinfo{year}{2009}), ISSN
  \bibinfo{issn}{0165-1684}, \bibinfo{note}{special Section: Visual Information
  Analysis for Security}.

\bibitem[{\citenamefont{Rudnicki et~al.}(2012)\citenamefont{Rudnicki, Walborn,
  and Toscano}}]{rudnicki12b}
\bibinfo{author}{\bibfnamefont{{\L}.}~\bibnamefont{Rudnicki}},
  \bibinfo{author}{\bibfnamefont{S.~P.} \bibnamefont{Walborn}},
  \bibnamefont{and} \bibinfo{author}{\bibfnamefont{F.}~\bibnamefont{Toscano}},
  \bibinfo{journal}{Phys. Rev. A} \textbf{\bibinfo{volume}{85}},
  \bibinfo{pages}{042115} (\bibinfo{year}{2012}).

\bibitem[{\citenamefont{Rudnicki}(2015)}]{rudnicki15}
\bibinfo{author}{\bibfnamefont{{\L}.}~\bibnamefont{Rudnicki}},
  \bibinfo{journal}{Phys. Rev. A} \textbf{\bibinfo{volume}{91}},
  \bibinfo{pages}{032123} (\bibinfo{year}{2015}).

\bibitem[{\citenamefont{Partovi}(1983)}]{partovi83}
\bibinfo{author}{\bibfnamefont{M.~H.} \bibnamefont{Partovi}},
  \bibinfo{journal}{Phys. Rev. Lett.} \textbf{\bibinfo{volume}{50}},
  \bibinfo{pages}{1883} (\bibinfo{year}{1983}).

\bibitem[{\citenamefont{{{\L}. Rudnicki} et~al.}(2012)\citenamefont{{{\L}.
  Rudnicki}, {S. P. Walborn}, and {F. Toscano}}}]{rudnicki12a}
\bibinfo{author}{\bibnamefont{{{\L}. Rudnicki}}},
  \bibinfo{author}{\bibnamefont{{S. P. Walborn}}}, \bibnamefont{and}
  \bibinfo{author}{\bibnamefont{{F. Toscano}}}, \bibinfo{journal}{EPL}
  \textbf{\bibinfo{volume}{97}}, \bibinfo{pages}{38003} (\bibinfo{year}{2012}).

\bibitem[{\citenamefont{Ray and van Enk}(2013{\natexlab{a}})}]{ray13a}
\bibinfo{author}{\bibfnamefont{M.~R.} \bibnamefont{Ray}} \bibnamefont{and}
  \bibinfo{author}{\bibfnamefont{S.~J.} \bibnamefont{van Enk}},
  \bibinfo{journal}{Phys. Rev. A} \textbf{\bibinfo{volume}{88}},
  \bibinfo{pages}{042326} (\bibinfo{year}{2013}{\natexlab{a}}),
  \urlprefix\url{https://link.aps.org/doi/10.1103/PhysRevA.88.042326}.

\bibitem[{\citenamefont{Ray and van Enk}(2013{\natexlab{b}})}]{ray13b}
\bibinfo{author}{\bibfnamefont{M.~R.} \bibnamefont{Ray}} \bibnamefont{and}
  \bibinfo{author}{\bibfnamefont{S.~J.} \bibnamefont{van Enk}},
  \bibinfo{journal}{Phys. Rev. A} \textbf{\bibinfo{volume}{88}},
  \bibinfo{pages}{062327} (\bibinfo{year}{2013}{\natexlab{b}}),
  \urlprefix\url{https://link.aps.org/doi/10.1103/PhysRevA.88.062327}.

\bibitem[{\citenamefont{Tasca et~al.}(2013)\citenamefont{Tasca, Rudnicki,
  Gomes, Toscano, and Walborn}}]{tasca13}
\bibinfo{author}{\bibfnamefont{D.~S.} \bibnamefont{Tasca}},
  \bibinfo{author}{\bibfnamefont{{\L}.}~\bibnamefont{Rudnicki}},
  \bibinfo{author}{\bibfnamefont{R.~M.} \bibnamefont{Gomes}},
  \bibinfo{author}{\bibfnamefont{F.}~\bibnamefont{Toscano}}, \bibnamefont{and}
  \bibinfo{author}{\bibfnamefont{S.~P.} \bibnamefont{Walborn}},
  \bibinfo{journal}{Phys. Rev. Lett.} \textbf{\bibinfo{volume}{110}},
  \bibinfo{pages}{210502} (\bibinfo{year}{2013}).

\bibitem[{\citenamefont{Gilchrist et~al.}(1998)\citenamefont{Gilchrist, Deuar,
  and Reid}}]{gilchrist98}
\bibinfo{author}{\bibfnamefont{A.}~\bibnamefont{Gilchrist}},
  \bibinfo{author}{\bibfnamefont{P.}~\bibnamefont{Deuar}}, \bibnamefont{and}
  \bibinfo{author}{\bibfnamefont{M.~D.} \bibnamefont{Reid}},
  \bibinfo{journal}{Phys. Rev. Lett.} \textbf{\bibinfo{volume}{80}},
  \bibinfo{pages}{3169} (\bibinfo{year}{1998}),
  \urlprefix\url{https://link.aps.org/doi/10.1103/PhysRevLett.80.3169}.

\bibitem[{\citenamefont{Banaszek and W\'odkiewicz}(1999)}]{banaszek99}
\bibinfo{author}{\bibfnamefont{K.}~\bibnamefont{Banaszek}} \bibnamefont{and}
  \bibinfo{author}{\bibfnamefont{K.}~\bibnamefont{W\'odkiewicz}},
  \bibinfo{journal}{Phys. Rev. Lett.} \textbf{\bibinfo{volume}{82}},
  \bibinfo{pages}{2009} (\bibinfo{year}{1999}),
  \urlprefix\url{https://link.aps.org/doi/10.1103/PhysRevLett.82.2009}.

\bibitem[{\citenamefont{Wenger et~al.}(2003)\citenamefont{Wenger, Hafezi,
  Grosshans, Tualle-Brouri, and Grangier}}]{wenger03}
\bibinfo{author}{\bibfnamefont{J.}~\bibnamefont{Wenger}},
  \bibinfo{author}{\bibfnamefont{M.}~\bibnamefont{Hafezi}},
  \bibinfo{author}{\bibfnamefont{F.}~\bibnamefont{Grosshans}},
  \bibinfo{author}{\bibfnamefont{R.}~\bibnamefont{Tualle-Brouri}},
  \bibnamefont{and} \bibinfo{author}{\bibfnamefont{P.}~\bibnamefont{Grangier}},
  \bibinfo{journal}{Phys. Rev. A} \textbf{\bibinfo{volume}{67}},
  \bibinfo{pages}{012105} (\bibinfo{year}{2003}),
  \urlprefix\url{https://link.aps.org/doi/10.1103/PhysRevA.67.012105}.

\bibitem[{\citenamefont{Vernaz-Gris et~al.}(2014)\citenamefont{Vernaz-Gris,
  Ketterer, Keller, Walborn, Coudreau, and Milman}}]{vernazgris14}
\bibinfo{author}{\bibfnamefont{P.}~\bibnamefont{Vernaz-Gris}},
  \bibinfo{author}{\bibfnamefont{A.}~\bibnamefont{Ketterer}},
  \bibinfo{author}{\bibfnamefont{A.}~\bibnamefont{Keller}},
  \bibinfo{author}{\bibfnamefont{S.~P.} \bibnamefont{Walborn}},
  \bibinfo{author}{\bibfnamefont{T.}~\bibnamefont{Coudreau}}, \bibnamefont{and}
  \bibinfo{author}{\bibfnamefont{P.}~\bibnamefont{Milman}},
  \bibinfo{journal}{Phys. Rev. A} \textbf{\bibinfo{volume}{89}},
  \bibinfo{pages}{052311} (\bibinfo{year}{2014}).

\bibitem[{\citenamefont{Ketterer et~al.}(2016)\citenamefont{Ketterer, Keller,
  Walborn, Coudreau, and Milman}}]{ketterer16}
\bibinfo{author}{\bibfnamefont{A.}~\bibnamefont{Ketterer}},
  \bibinfo{author}{\bibfnamefont{A.}~\bibnamefont{Keller}},
  \bibinfo{author}{\bibfnamefont{S.~P.} \bibnamefont{Walborn}},
  \bibinfo{author}{\bibfnamefont{T.}~\bibnamefont{Coudreau}}, \bibnamefont{and}
  \bibinfo{author}{\bibfnamefont{P.}~\bibnamefont{Milman}},
  \bibinfo{journal}{Phys. Rev. A} \textbf{\bibinfo{volume}{94}},
  \bibinfo{pages}{022325} (\bibinfo{year}{2016}).

\bibitem[{\citenamefont{Laversanne-Finot
  et~al.}(2017)\citenamefont{Laversanne-Finot, Ketterer, Barros, Walborn,
  Coudreau, Keller, and Milman}}]{finot17}
\bibinfo{author}{\bibfnamefont{A.}~\bibnamefont{Laversanne-Finot}},
  \bibinfo{author}{\bibfnamefont{A.}~\bibnamefont{Ketterer}},
  \bibinfo{author}{\bibfnamefont{M.~R.} \bibnamefont{Barros}},
  \bibinfo{author}{\bibfnamefont{S.~P.} \bibnamefont{Walborn}},
  \bibinfo{author}{\bibfnamefont{T.}~\bibnamefont{Coudreau}},
  \bibinfo{author}{\bibfnamefont{A.}~\bibnamefont{Keller}}, \bibnamefont{and}
  \bibinfo{author}{\bibfnamefont{P.}~\bibnamefont{Milman}},
  \bibinfo{journal}{Journal of Physics A: Mathematical and Theoretical}
  \textbf{\bibinfo{volume}{50}}, \bibinfo{pages}{155304}
  (\bibinfo{year}{2017}).

\bibitem[{\citenamefont{Paul et~al.}(2018)\citenamefont{Paul, Walborn, Tasca,
  and Rudnicki}}]{paul18}
\bibinfo{author}{\bibfnamefont{E.~C.} \bibnamefont{Paul}},
  \bibinfo{author}{\bibfnamefont{S.~P.} \bibnamefont{Walborn}},
  \bibinfo{author}{\bibfnamefont{D.~S.} \bibnamefont{Tasca}}, \bibnamefont{and}
  \bibinfo{author}{\bibfnamefont{{\L}.}~\bibnamefont{Rudnicki}},
  \bibinfo{journal}{Phys. Rev. A} \textbf{\bibinfo{volume}{97}},
  \bibinfo{pages}{052103} (\bibinfo{year}{2018}),
  \urlprefix\url{https://link.aps.org/doi/10.1103/PhysRevA.97.052103}.

\bibitem[{\citenamefont{Silva et~al.}(2020)\citenamefont{Silva, Rudnicki,
  Tasca, and Walborn}}]{silva21}
\bibinfo{author}{\bibfnamefont{T.~L.} \bibnamefont{Silva}},
  \bibinfo{author}{\bibfnamefont{{\L}.}~\bibnamefont{Rudnicki}},
  \bibinfo{author}{\bibfnamefont{D.~S.} \bibnamefont{Tasca}}, \bibnamefont{and}
  \bibinfo{author}{\bibfnamefont{S.~P.} \bibnamefont{Walborn}},
  \emph{\bibinfo{title}{Periodic discretized continuous observables are neither
  continuous nor discrete}} (\bibinfo{year}{2020}),
  \eprint{quant-ph/2009.05062}.

\bibitem[{\citenamefont{Tasca et~al.}(2018{\natexlab{b}})\citenamefont{Tasca,
  Rudnicki, Aspden, Padgett, Souto~Ribeiro, and Walborn}}]{tasca18b}
\bibinfo{author}{\bibfnamefont{D.~S.} \bibnamefont{Tasca}},
  \bibinfo{author}{\bibfnamefont{{\L}.}~\bibnamefont{Rudnicki}},
  \bibinfo{author}{\bibfnamefont{R.~S.} \bibnamefont{Aspden}},
  \bibinfo{author}{\bibfnamefont{M.~J.} \bibnamefont{Padgett}},
  \bibinfo{author}{\bibfnamefont{P.~H.} \bibnamefont{Souto~Ribeiro}},
  \bibnamefont{and} \bibinfo{author}{\bibfnamefont{S.~P.}
  \bibnamefont{Walborn}}, \bibinfo{journal}{Phys. Rev. A}
  \textbf{\bibinfo{volume}{97}}, \bibinfo{pages}{042312}
  (\bibinfo{year}{2018}{\natexlab{b}}),
  \urlprefix\url{https://link.aps.org/doi/10.1103/PhysRevA.97.042312}.

\bibitem[{\citenamefont{Rudnicki}(2012)}]{rudnicki10}
\bibinfo{author}{\bibfnamefont{{\L}.}~\bibnamefont{Rudnicki}}, in
  \emph{\bibinfo{booktitle}{Proceedings of New Perspectives in Quantum
  Statistics and Correlations, M. Hiller, F. de Melo, P. Pickl, T. Wellens, S.
  Wimberger (Eds.)}} (\bibinfo{publisher}{Universitatsverlag Winter},
  \bibinfo{year}{2012}), p.~\bibinfo{pages}{49}, \eprint{arXiv:1010.3269}.

\bibitem[{\citenamefont{Namias}(1980)}]{FFT}
\bibinfo{author}{\bibfnamefont{V.}~\bibnamefont{Namias}}, \bibinfo{journal}{IMA
  Journal of Applied Mathematics} \textbf{\bibinfo{volume}{25}},
  \bibinfo{pages}{241} (\bibinfo{year}{1980}).

\bibitem[{\citenamefont{Weigert and Wilkinson}(2008)}]{weigert08}
\bibinfo{author}{\bibfnamefont{S.}~\bibnamefont{Weigert}} \bibnamefont{and}
  \bibinfo{author}{\bibfnamefont{M.}~\bibnamefont{Wilkinson}},
  \bibinfo{journal}{Phys. Rev. A} \textbf{\bibinfo{volume}{78}},
  \bibinfo{pages}{020303} (\bibinfo{year}{2008}).

\bibitem[{\citenamefont{S{\'a}nchez}(1993)}]{SANCHEZ1993233}
\bibinfo{author}{\bibfnamefont{J.}~\bibnamefont{S{\'a}nchez}},
  \bibinfo{journal}{Physics Letters A} \textbf{\bibinfo{volume}{173}},
  \bibinfo{pages}{233} (\bibinfo{year}{1993}), ISSN \bibinfo{issn}{0375-9601},
  \urlprefix\url{https://www.sciencedirect.com/science/article/pii/0375960193902696}.

\bibitem[{\citenamefont{Wu et~al.}(2009)\citenamefont{Wu, Yu, and
  M\o{}lmer}}]{PhysRevA.79.022104}
\bibinfo{author}{\bibfnamefont{S.}~\bibnamefont{Wu}},
  \bibinfo{author}{\bibfnamefont{S.}~\bibnamefont{Yu}}, \bibnamefont{and}
  \bibinfo{author}{\bibfnamefont{K.}~\bibnamefont{M\o{}lmer}},
  \bibinfo{journal}{Phys. Rev. A} \textbf{\bibinfo{volume}{79}},
  \bibinfo{pages}{022104} (\bibinfo{year}{2009}),
  \urlprefix\url{https://link.aps.org/doi/10.1103/PhysRevA.79.022104}.

\bibitem[{\citenamefont{Pucha\l{}a et~al.}(2015)\citenamefont{Pucha\l{}a,
  Rudnicki, Chabuda, Paraniak, and \ifmmode~\dot{Z}\else
  \.{Z}\fi{}yczkowski}}]{PhysRevA.92.032109}
\bibinfo{author}{\bibfnamefont{Z.}~\bibnamefont{Pucha\l{}a}},
  \bibinfo{author}{\bibfnamefont{{\L}.}~\bibnamefont{Rudnicki}},
  \bibinfo{author}{\bibfnamefont{K.}~\bibnamefont{Chabuda}},
  \bibinfo{author}{\bibfnamefont{M.}~\bibnamefont{Paraniak}}, \bibnamefont{and}
  \bibinfo{author}{\bibfnamefont{K.}~\bibnamefont{\ifmmode~\dot{Z}\else
  \.{Z}\fi{}yczkowski}}, \bibinfo{journal}{Phys. Rev. A}
  \textbf{\bibinfo{volume}{92}}, \bibinfo{pages}{032109}
  (\bibinfo{year}{2015}),
  \urlprefix\url{https://link.aps.org/doi/10.1103/PhysRevA.92.032109}.

\bibitem[{\citenamefont{Wiseman et~al.}(2007)\citenamefont{Wiseman, Jones, and
  Doherty}}]{wiseman07}
\bibinfo{author}{\bibfnamefont{H.~M.} \bibnamefont{Wiseman}},
  \bibinfo{author}{\bibfnamefont{S.~J.} \bibnamefont{Jones}}, \bibnamefont{and}
  \bibinfo{author}{\bibfnamefont{A.~C.} \bibnamefont{Doherty}},
  \bibinfo{journal}{Phys. Rev. Lett.} \textbf{\bibinfo{volume}{98}},
  \bibinfo{eid}{140402} (\bibinfo{year}{2007}).

\end{thebibliography}

\end{document}